\documentclass[epsfig,indentfirst,bm,12pt]{article}
\usepackage{CJK}
\usepackage{graphicx,amsmath,dcolumn,bm}
\usepackage{color}
\begin{document}
\title{\Large{\bf{The study of gluon energy loss in cold nuclear matter from $J/\psi$ production}}
\thanks {Supported partially by National Natural Science
Foundation of China(11405043) and Natural Science Foundation of Hebei Province(A2018209269).}}

\begin{CJK*}{GBK}{song}
\author{Li-Hua Song $^{1}$
\footnote{\tt{ E-mail:songlh@ncst.edu.cn}}
Lin-Wan Yan $^{1}$
Shang-Fei Xin $^{1}$}

\date{}

\maketitle

\end{CJK*}

\noindent {\small 1.College of
Science,  North China University of Science and Technology, Tangshan 063210,
China}

\baselineskip 9mm
\begin{abstract}
The energy loss effects of the incident quark, gluon and the color octet $c\bar{c}$ on $J/\psi$ suppression in p-A collisions are studied by means of the experimental data at E866, RHIC and LHC energy. By the Salgado-Wiedemann (SW) quenching weights and the recent EPPS16 nuclear parton distribution functions together with nCTEQ15, we extract the transport coefficient for the gluon energy loss from the E866 experimental data in the middle $x_{F}$ region ($0.20<x_{F}<0.65$). It is found that the difference between the values of the transport coefficient for light quark, gluon and heavy quark in the cold nuclear matter is very small.
 The theoretical results modified by the parton energy loss effects are consistent with the experimental data for E866 and RHIC energy, and the gluon energy loss plays a remarkable role on $J/\psi$ suppression in a broad variable range. Since the corrections of the nuclear parton distribution functions in the $J/\psi$ channel are significant at LHC energy, the nuclear modification due to the parton energy loss is minimal. It is worth noting that we use CEM at leading order to compute the p-p baseline, and the conclusion in this paper is CEM model dependent.

\vskip 1.0cm



\end{abstract}

\maketitle
\newpage
\vskip 0.5cm

\section{Introduction}
In the past thirty years, the study of the origin of $J/\psi$ suppression in p-A collisions has been important for the understanding of the Quark-Gluon Plasma (QGP, the state of matter where quarks and gluons are deconfined). Due to the Debye color screening of the charm-quark potential, the high quark density in the QGP formation would suppress the yield of $J/\psi$ mesons in high-energy heavy-ion collisions relative to that in p-p collisions[1]. The generic features of hard QCD processes in the nuclear environment can be quantified from the experimental data of $J/\psi$ production in p-A collisions, which can help to perform the reliable baseline predictions about the properties of the QGP formed in heavy-ion collisions.

In $J/\psi$ production from p-A collisions, the charmonium production can be modified by cold nuclear
matter effects, such as the parton energy loss [2], the nuclear absorption and the modification of
the parton distribution functions due to the nuclear environment[3]. It is necessary to constrain the so-called cold
nuclear matter effects on quarkonium production, in order to interpret these results unambiguously in heavy-ion collisions.
The wealth of the data available in p-A collisions from a wide collision energy range ( e.g., NA3[4], E772[5], E866[6,7], NA50[8], HEAR-B[9], LHC[10,11] and RHIC[12]) provides the good probes of the origin of $J/\psi$ suppression. However, the conventional nuclear suppression mechanism of  $J/\psi$ suppression is still an open question, due to the presence of various competing effects, depending on the precise kinematics and collision energy. Up to date,  some of the phenomenological approaches are proposed on the basis of hadron nuclear absorption or the shadowing of the gluon distribution for the target nucleus expected in the small $x_{2}$ range ($x_{2}<10^{-2}$), and another fundamental phenomenological model assumes that the parton energy loss induced by parton multiple scattering in the nuclear environment plays a decisive role in $J/\psi$ suppression[13].

Since the intensity of each cold nuclear matter effect is usually unknown a priori, it is a sound strategy to investigate each of these effects separately by comparing all available data systematically and quantitatively, while maintaining the minimum number of the assumptions and free parameters. We find it is appealing that the modification of
the parton distribution functions and parton energy loss effect can describe the $J/\psi$ suppression measured in p-A collisions when a high-energy $J/\psi$  is formed long after the nucleus. Consequently, following our previous work about the parton energy loss effect in cold nuclear matter[14-16], in this paper, we present the phenomenological approach with assuming that the $J/\psi$ suppression observed is mainly induced by the nuclear PDF effects and the energy loss effects of the incoming quark, gluon and the color octet $c\bar{c}$. By the Salgado-Wiedemann (SW) quenching weights[17], we will investigate the incident quark energy loss, the incident gluon energy loss and the color octet $c\bar{c}$ energy loss separately by comparing the E866[6,7], LHC[10,11] and RHIC[12] experimental data. Since the strength of each parton energy loss depends on a single free parameter named the transport coefficient which characterizes the medium-induced transverse momentum squared transferred to the projectile per unit path length, this research would help to understand the microscopic dynamics of
medium-induced parton energy loss corresponding to the color charge of the
parton.

The arrangement of this article is as follows. The theoretical framework of our study is introduced
in Section II, and the results for the nuclear modification of $J/\psi$ production on the basis of the nuclear PDF effects and parton energy loss effect are given in Sections III. We then summarize our main conclusions.

\section{$J/\psi$ production modified due to energy loss effect}
In $J/\psi$ production from p-A collisions, a hard parton traveling through the nuclear target undergoes the multiple soft collisions which induce gluon emission. Medium-induced gluon radiation modifies the correspondence
between the initial parton and the final hadron momenta. With assuming that these radiated gluons take away an energy $\varepsilon$ from the leading particle, this modification is determined by the probability distribution $D(\varepsilon)$ in the energy loss. If gluons are emitted independently,
$D(\varepsilon)$ is the normalized sum of the emission probabilities
for an arbitrary number of $n$ gluons which carry away the
total energy $\varepsilon$[17]:
\begin{eqnarray}
D(\varepsilon)=\sum_{n=0}^{\infty}\frac{1}{n!}[\prod_{i=1}^{n}\int d\omega_{i}\frac{dI(\omega_{i})}{d\omega}]\delta(\varepsilon-\sum_{i=1}^{n}\omega_{i})exp[-\int_{0}^{+\infty}d\omega\frac{dI(\omega)}{d\omega}].
\end{eqnarray}
Here, $dI(\omega)/d\omega$ is the medium-induced gluon spectrum.
According to Ref.[18], the probability distribution $D(\varepsilon)$ has
a discrete and a continuous part:
\begin{eqnarray}
D(\varepsilon)=p_{0}\delta(\varepsilon)+p(\varepsilon).
\end{eqnarray}
Its normalization is unity.
In this paper, the probability distribution $D(\varepsilon)$ used are calculated in the multiple soft and
single hard scattering approximations, and the results named as SW quenching weights are available as a FORTRAN routine[19].
The SW quenching weight is a scaling function of the variable $\varepsilon/E$ and $\varepsilon/\omega_{c}$, with $E$ representing the parton energy after radiating the energy  $\varepsilon$ and $\omega_{c}=\frac{1}{2}\hat{q}L^{2}$.  Here, $\hat{q}$ denotes the transport coefficient and $L
$ is the path length covered by the parton in the nuclear medium. Generally, the cross section in the medium can  be modified as:
\begin{eqnarray}
\sigma^{medium}=D(\varepsilon)\bigotimes\sigma^{vacuum}.
\end{eqnarray}
Then, the charmonium production cross section $d\sigma_{p-A}/dx_{F}$ can be expressed as:
\begin{eqnarray}
\frac{{d\sigma}_{p-A}}{{dx_{F}}}(x_{F})&=&\int_{0}^{\varepsilon_{max}}d\varepsilon
D(\varepsilon)\frac{{d\sigma}_{p-p}}{{dx_{F}}}(\varepsilon,x_{F}),
\end{eqnarray}
where the upper limit on the energy loss is $\varepsilon_{max}=min(E_{p}-E, E)$ with $E_{p}$  denoting the beam  energy in the rest frame of the target nucleus.

This medium-induced energy loss leads to the rescaling of the parton momentum fraction. In view of the energy loss effect of the color octet $c\bar{c}$ pair, the observed $J/\psi$ at a given
$x_{F}$ actually comes from a $c\bar{c}$ pair originally produced at the higher value $x'_{F}=x_{F}+\Delta x_{F}$ with $\Delta
x_{F}=\varepsilon_{c\bar{c}}/E_{p}$. The energy loss of the incoming gluon (quark) also results in a change in its momentum fraction prior to the collision: $\Delta x_{1g}=\varepsilon_{g}/E_{p}$ ($\Delta x_{1q}=\varepsilon_{q}/E_{p}$). In order to incorporate the features of the process for $J/\psi$ production, there are two acceptable formalisms have been presented, the non-relativistic QCD (NRQCD) [20] and the color evaporation
model (CEM)[21], which have been successful in charmonium phenomenology. In this work, we choose the CEM, as the CEM has fewer free parameters[22-26].
 According to the CEM[21] and considering the above rescalings, the cross section $d\sigma_{p-p}/dx_{F}$ ( a convolution
of the $q\overline{q}$ cross section $(\sigma_{q\bar{q}})$ and $gg$ cross section ($\sigma_{gg}$) with the parton distribution
functions $f_{i}$ in the incident proton and $f^{A}_{i}$ in the target proton) can be expressed as:
\begin{eqnarray}
\frac{{d\sigma}_{p-p}}{{dx_{F}}}(x_{F},\varepsilon_{c\bar{c},g,q})=\rho_{J/\psi}\int^{2m_{D}}_{2m_{c}}dm\frac{2m}{\sqrt{x_{F}^{2}s+4m^{2}}}
\times[f_{g}(x'_{1g},m^{2})f^{A}_{g}(x'_{2g},m^{2})\sigma_{gg}(m^{2})\nonumber +\\
 \sum^{}_{q=u,d,s}\{f_{q}(x'_{1q},m^{2})f^{A}_{\bar{q}}(x'_{2q},m^{2})
+{f_{\bar{q}}(x'_{1q},m^{2})f^{A}_{q}(x'_{2q},m^{2})}
\}\sigma_{q\bar{q}}(m^{2})],
\end{eqnarray}
with $x'_{1g(q)}=x'_{1}+\Delta x_{1g(q)}$,
 $x'_{1}=\frac{1}{2}[\sqrt{(x'_{F})^{2}(1-m^{2}/s)^{2}+4m^{2}/s}+x'_{F}(1-m^{2}/s)]$, $m^{2}=x_{1}x_{2}s$ in the rest frame of the target nuclei, and $\rho_{J/\psi}$ denoting the fraction of the $c\overline{c}$ pair evolving into the $J/\psi$ state. It is mentioned that the following calculation is based on leading order in CEM, especially on the solutions of momentum fractions.

Analogously, due to the energy loss effects, the charmonium production cross section as a function
of the rapidity $y$ (with $x_{1(2)}=\frac{m}{\sqrt{s}}e^{\pm y}$ and $E=E_{p}e^{y}m/\sqrt{s}$ ) can be modified as:
\begin{eqnarray}
\frac{{d\sigma}_{p-p}}{{dy}}(y)&=&\rho_{J/\psi}\int^{2m_{D}}_{2m_{c}}dm\frac{2m}{s}
\times[f_{g}(x'_{1g},m^{2})f^{A}_{g}(x'_{2g},m^{2})\sigma_{gg}(m^{2})\nonumber \\
& & +\sum^{}_{q=u,d,s}\{f_{q}(x'_{1q},m^{2})f^{A}_{\bar{q}}(x'_{2q},m^{2})
+{f_{\bar{q}}(x'_{1q},m^{2})f^{A}_{q}(x'_{2q},m^{2})}
\}\sigma_{q\bar{q}}(m^{2})].
\end{eqnarray}
Here, $x'_{1g(q)}=\frac{m}{\sqrt{s}}e^{y'}+\Delta x_{1g(q)}$ with $y'=y+ln(\frac{E+\varepsilon_{c\bar{c}}}{E})$ .

\section{Results and discussion}
The amount of medium-induced gluon radiation and the strength of the $J/\psi$ suppression in p-A collisions are controlled by the transport coefficient $\hat{q}$ in the target nucleus. Based on the above parton energy loss model expressed in section 2, we extract the transport coefficient $\hat{q}_{g}$ for the gluon energy loss from the E866 experimental data[6,7] by means of the SW quenching weights for gluons[17]. It is worth noting that according to Ref.[27] the $c\bar{c}$ remains colored on its entire path for the E866 energy ( $\sqrt{s}=38.7$ GeV ) in the range $0.20\leq x_{F} \leq 0.65 $. The extracted results of the transport coefficient $\hat{q}_{g}$ are summarized in Table I, by using the CERN subroutine MINUIT[28] and the recent EPPS16 nuclear parton distributions [29] together with nCTEQ15 parton density in the proton bound in a nucleus[30]. In this calculation, we employ the values of the path $L$ obtained from the Glauber model calculation using realistic nuclear densities[31], $\hat{q}_{q}=0.32\pm0.04$ $GeV^{2}/fm$ extracted from the nuclear Drell-Yan experimental data by means of the SW quenching weights for light quarks [15] and $\hat{q}_{c\bar{c}}=0.29\pm0.07$ $GeV^{2}/fm$ obtained by the SW quenching weights for heavy quarks[32].

From Table I, it is shown that the theoretical results are in good agreement with the E866 experimental data in $0.20<x_{F}<0.65$ range ($\chi^{2}/ndf=1.10$). The experimental data including the small and large $x_{F}$ region deviate from the calculated results ($\chi^{2}/ndf=23.19$), which illustrates that the role of other nuclear effects (such as gluon saturation at small $x_{F}$ region and nuclear absorption at big $x_{F}$ region) on the modification of the charmonium production cross section can not be ignored in the small or big $x_{F}$ range. Consequently, we determine the transport coefficient $\hat{q}_{g}$ ($\hat{q}_{g}=0.31\pm0.02$ $GeV^{2}/fm$) for the gluon energy loss from the data in the middle region $0.20<x_{F}<0.65$. It is found that the difference between these values of the transport coefficient for light quark, gluon and heavy quark in the cold nuclear matter is very small. This conclusion is contradict with the well known statement that the transport coefficient $\hat{q}_{g}$ is larger than that of quark due to the ratio of the Casimir factors $C_{A}/C_{F}=9/4$ in leading logarithmic approximation. It is worth emphasizing that the obtained energy loss of an incoming quark from Drell-Yan experimental data depends strongly on the nuclear parton distribution functions (see Ref. [33] for more detail discussion). The recent EPPS16 nuclear parton distributions provide the uncertainty estimates are more objective flavor by flavor[29]. The errors of $\hat{q}_{g}$ from the uncertainty of nuclear PDFs EPPS16 are analyzed. The extracted value of $\hat{q}_{g}$ to the specific error sets $S_{1}^{-}$ ($S_{1}^{+}$)for EPPS16 is $\hat{q}_{g}=0.30\pm0.02$ $GeV^{2}/fm$ ($\hat{q}_{g}=0.32\pm0.02$ $GeV^{2}/fm$). In addition, the nuclear modification for gluon distribution function is apparently different between the different sets, as discussed in Ref. [34]. The  uncertainty of the gluon distribution function may be the main source of the uncertainty associated with the result of the transport coefficient $\hat{q}_{g}$.In order to further check the initial state PDFs dependence in determining the jet transport coefficient, we simply eliminate the nuclear dependence of PDFs, and by using nCTEQ15 parton density the transport coefficient $\hat{q}_{g}=0.35\pm0.03$ $GeV^{2}/fm$ extracted from the E866 experimental data for the range $0.20 < x_{F} < 0.65$.
\begin{table}[tb]
\caption{The values of $\hat{q}_{g}$ and $\chi^{2}/ndf$ extracted from the E866 experimental data[6,7].}
\begin{center}
\begin{tabular}{p{5cm}p{3cm}p{3cm}p{3cm}c}\hline

$x_{F}$& No.data   & $\hat{q}_{g}$ (GeV$^{2}$/fm) & $\chi^{2}/ndf$  \\
\hline
0.20$<x_{F}<$0.65&18&0.31$\pm$0.02&1.10\\
0.30$<x_{F}<$0.95&26&0.26$\pm$0.01&15.41\\
0.20$<x_{F}<$0.95&44&0.26$\pm$0.01&10.29\\
0.00$<x_{F}<$0.95&49&0.25$\pm$0.01&23.19\\

\hline
\end{tabular}
\end{center}
\end{table}

\begin{figure}[tb]
\centering
\includegraphics*[width=16.5cm, height=13cm]{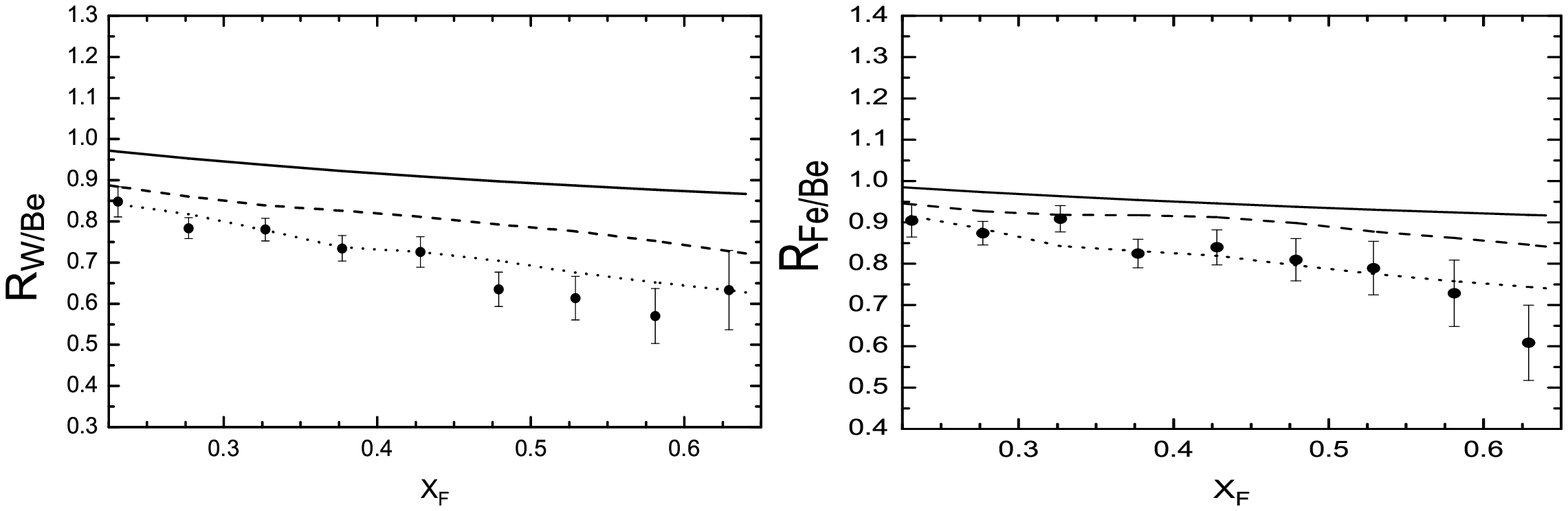}
\vspace{-8.0cm} \caption{The $J/\psi$ production cross section ratios $R_{W(Fe)/Be}(x_{F})$ for $0.20<x_{F}<0.65$ . The solid lines denote the results modified only by EPPS16 nuclear parton distributions, the dashed curves represent the calculations
including the incident quark energy loss in the initial state and the $c\bar{c}$ energy loss in the final state,  and the results together with the incident gluon energy loss correction are shown as the dotted lines. The experimental points are taken from
the E866 data for $J/\psi$ production in p-A collisions [6,7]. }
\end{figure}
\begin{figure}[tb]
\centering
\includegraphics*[width=14cm, height=11.5cm]{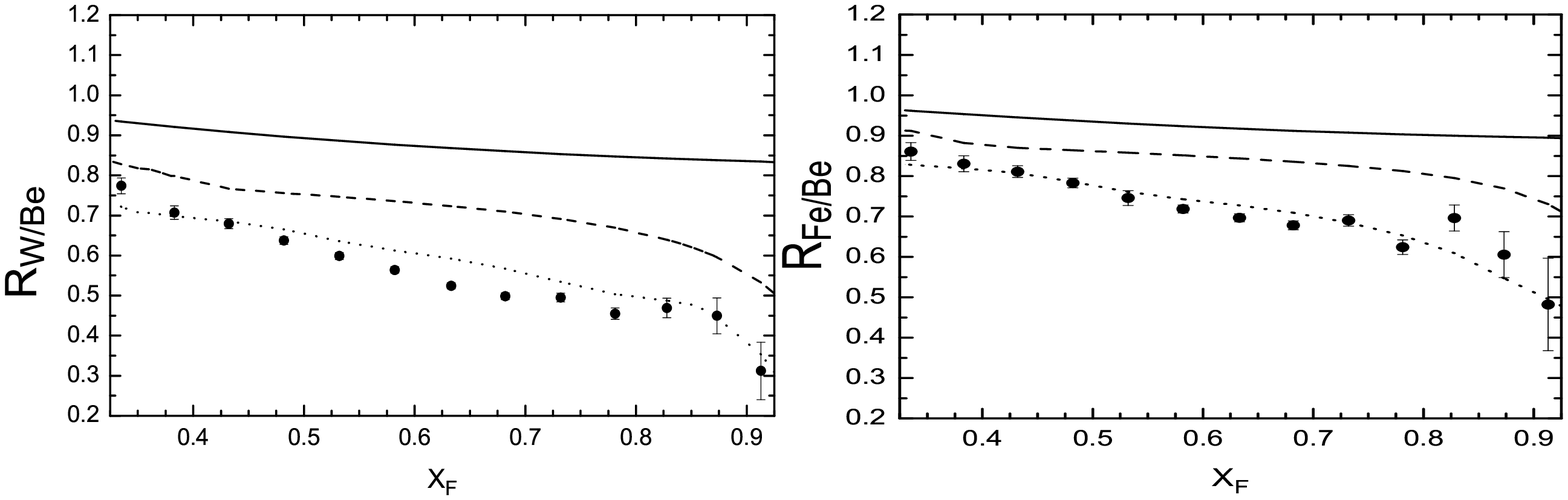}
\vspace{-6cm} \caption{The $J/\psi$ production cross section ratios $R_{W(Fe)/Be}(x_{F})$ for $0.3<x_{F}<0.95$. The dashed curves represent the calculations with the $c\bar{c}$ energy loss. The other comments are the same as those in Fig. 1.}
\end{figure}

In Fig.1, with $\hat{q}_{g}=0.31\pm0.02$ $GeV^{2}/fm$, $\hat{q}_{q}=0.32\pm0.04$ $GeV^{2}/fm$ and $\hat{q}_{c\bar{c}}=0.29\pm0.07$ $GeV^{2}/fm$, we firstly compare the model predictions with the E866 data for $J/\psi$ production in the range $0.20<x_{F}<0.65$. The solid lines denote the results modified only by EPPS16 nuclear parton distributions, the dashed curves represent the calculations including the incident quark energy loss in the initial state and the $c\bar{c}$ energy loss in the final state, and the results together with the incident gluon energy loss correction are shown as the dotted lines. The good agreement reported in the dotted lines from Fig.1 fully supports this energy loss model with the energy loss probability distribution calculated by SW quenching weights. From Fig.1, it is also remarkable that the energy loss of incident gluon in the initial state plays a certain role on $J/\psi$ suppression. This indicates that the E866 data for $J/\psi$ production in the range $0.20<x_{F}<0.65$ would provide a good way to discriminatingly identify the gluon energy loss. In Fig.2, the calculated results are also compared with the E866 data for $0.30<x_{F}<0.95$. The strength of the $J/\psi$ suppression induced by the $c\bar{c}$ energy loss effect are shown as the dashed lines in Fig.2, which indicates that   the nuclear modification due to the $c\bar{c}$ energy loss are more significant at larger targets for larger $x_{F}$ region. As can be also seen in Fig.2, the predictions of the dotted lines (including the correction induced by the quark, gluon and $c\bar{c}$ energy loss) are consistent with the experimental data for $0.30<x_{F}<0.95$. From the solid lines in Fig.1 and Fig.2, we can see that the nuclear suppression from the nuclear modification of EPPS16 nuclear parton distributions becomes larger as the increase of $x_{F}$ in the range $0.20<x_{F}<0.95$. The nucleon parton momentum fractions $x_{2}$ is $1.84\times 10^{-2}$ to $4.22\times 10^{-3}$ accordance to the $x_{F}$ range from $0.20$ to $0.95$. In this $ x_{2}$ range, the shadowing effect plays the main role. The suppression induced by the shadowing effect of the nuclear parton distributions is larger with the decrease of $x_{2}$. The tendency of the dashed and dotted lines in Fig.1 and Fig.2 indicates that the suppression due to the $c\bar{c}$ and incoming gluon energy loss also becomes larger as the increase of $x_{F}$ in the range $0.20<x_{F}<0.95$.

\begin{figure}[tb]
\centering
\includegraphics*[width=15cm, height=10cm]{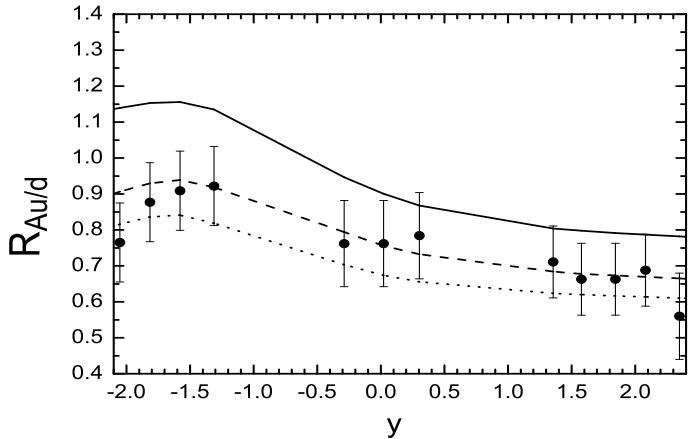}
\vspace{-3.5cm} \caption{The $J/\psi$ production cross section ratios $R_{Au/d}(y)$.  The experimental points are taken from
 RHIC experimental data [12]. The other comments are the same as those in Fig. 1.}
\end{figure}

Furthermore, we discuss the $J/\psi$ suppression expected in d-Au collisions at the RHIC ($\sqrt{s}$ = 200 GeV)[12] and in p-Pb collisions at the LHC ($\sqrt{s}$ = 5.0 TeV)[10,11] . The comparisons between RHIC data and our model predictions are shown in Fig.3. The solid lines denote the results modified only by EPPS16 nuclear parton distributions, the dashed curves represent the calculations including the incident quark energy loss in the initial state and the $c\bar{c}$ energy loss in the final state, and the results together with the incident gluon energy loss correction are shown as the dotted lines. As can be seen from Fig.3, the energy loss model can reproduce nicely the $J/\psi$ suppression in the whole variable interval. It is worth noting that a good agreement is also observed in some negative $y$ bins for $y < y^{crit}$ $(y^{crit}= -1.1)$, where nuclear absorption might also play a role. In addition, the prediction of the dotted line illustrates that the gluon energy loss can induce the obvious suppression of  $J/\psi$ production in a broad $y$ range. The tendency of the solid line modified only by the EPPS16 nuclear parton distributions in Fig.3 shows an enhancement in the range $-2.1<y<-1.5$, then falls with the increase of $y$ in the range $-1.5<y<2.4$. The physical reasons of $y$ dependence of the nuclear modification from EPPS16 densities are that the anti-shadowing effect induces the enhancement in the coverage $-2.0<y<-1.5$ ($0.09>x_{2}>0.05$), and shadowing effect causes the suppression for $-1.5<y<2.4$ ($0.05>x_{2}>0.001$). The parton energy loss effects have the analogous tendency as the increase of $y$ in the range $-2.1<y<2.4$.

In Fig.4, we show separately the $J/\psi$ production cross section ratios $R_{Pb/p}(y)$, modified only by EPPS16 nuclear parton distributions(solid lines), only by $c\overline{c}$ energy loss effect (dashed lines), only by the incident gluon energy loss (dotted lines), and only by the incident quark energy loss (dash dot lines). From Fig.4, we can see that the sole nuclear effects of the parton distributions might be responsible for the  $J/\psi$ observed suppression, and the nuclear modification due to the parton energy loss is minimal. This indicates that the corrections of the nuclear parton distribution functions in $J/\psi$ production are significant for the high center-of-mass energy of the collision at the LHC. In addition, we can see that the tendency of the nuclear modification of EPPS16 distributions decreases steeper with the increase of $y$ in the region $-2.0<y<-3.0$ and gradually in $-3.0<y<3.5$. For LHC energy, in the range $-2.0<y<3.5$ ($0.04>x_{2}>1.49\times10^{-5}$ and $5.47\times10^{-6}<x_{2}<1.63\times10^{-2}$), shadowing effect plays the main role.

\begin{figure}[tb]
\centering
\includegraphics*[width=15cm, height=10cm]{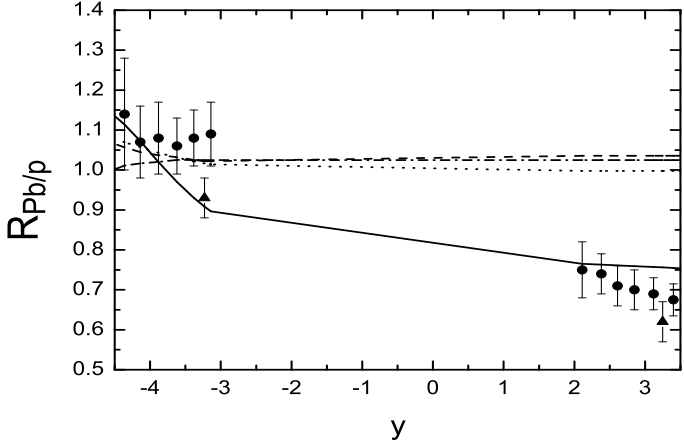}
\vspace{-3.5cm} \caption{The $J/\psi$ production
cross section ratios $R_{Pb/p}(y)$, modified only by EPPS16 nuclear parton distributions (solid lines), only by $c\overline{c}$ energy loss effect (dashed lines), only by the incident gluon energy loss (dotted lines), only by the incident quark energy loss (dash dot lines). The experimental data are separately form ALICE Collaboration [10] (filled circles) and LHCb Collaboration [11] (solid triangles).}
\end{figure}

\section{ Summary }
By the Salgado-Wiedemann (SW) quenching weights and the recent EPPS16 nuclear parton distribution functions together with nCTEQ15, we investigate respectively the energy loss effects of the incident quark, gluon and color octet $c\bar{c}$ on $J/\psi$ suppression in p-A collisions. The transport coefficient $\hat{q}_{g}$ ($\hat{q}_{g}=0.31\pm0.02$ $GeV^{2}/fm$, with $\chi^{2}/ndf=1.10$) for the gluon energy loss is extracted from the E866 data in the middle $x_{F}$ region $0.20<x_{F}<0.65$. This indicates that the difference between the values of the transport coefficient for light quark ($\hat{q}_{q}=0.32\pm0.04$ $GeV^{2}/fm$[15]), gluon
and heavy quark ($\hat{q}_{c\bar{c}}=0.29\pm0.07$ $GeV^{2}/fm$[25]) in the cold nuclear matter is very small. The good agreement between our model with the E866 and RHIC experimental data strongly supports that the parton energy loss is a dominant effect in p-A quarkonium nuclear suppression when $J/\psi$ formation occurring outside the nuclear target.  It is also found that the gluon energy loss in the initial state plays a remarkable role on $J/\psi$ suppression in a broad variable range at E866 and RHIC energy. However, the comparison between the $J/\psi$ suppression data at LHC energy and our theoretical predictions indicate that the effects of the nuclear parton distribution functions are significant for the high center-of-mass energy of the collision at the LHC. In the future work, it will be interesting to check whether the L-dependence of the energy loss
predicted in our present model is consistent with the centrality dependence of the LHC data. It is worth noting that we use CEM at leading order to compute the p-p baseline, and the conclusion in this paper is CEM model dependent. We hope this study about the parton energy loss effect on $J/\psi$ production from p-A collisions can give insight on the evaluation of $J/\psi$ suppression in A-A collisions.


\end{document}